\documentstyle[aps,prl,epsfig]{revtex}
\begin{document}
\draft
\newcommand{\pr}[3]{Phys. Rev. {\bf #1}, #2 (#3)}
\newcommand{\physb}[3]{Phys. Rev. B {\bf #1}, #2 (#3)}
\newcommand{\physl}[3]{Phys. Rev. Lett. {\bf #1}, #2 (#3)}
\twocolumn[\hsize\textwidth\columnwidth\hsize\csname
@twocolumnfalse\endcsname
\title{Fermi-Edge Singularities In Al$_x$Ga$_{1-x}$As Quantum Wells :\\
Extrinsic Versus Many-Body Scattering Processes}

\author{T. M\'elin\cite{byline} and F. Laruelle}

\address{Laboratoire de Microstructures et de
Micro\'electronique, Centre National de la Recherche Scientifique\\
196 Avenue Henri Ravera, BP 107, 92225 Bagneux Cedex, France}
\maketitle
\begin{abstract}
\noindent A Fano resonance mechanism is evidenced to
control the formation of optical Fermi-edge singularities in multi-subband systems
such as remotely doped Al$_x$Ga$_{1-x}$As heterostructures.
Using Fano parameters, we probe the {\it physical
nature }of the interaction
between Fermi sea electrons and empty conduction subbands.
We show that processes of extrinsic origin
like alloy-disorder prevail easily at 2D over
multiple diffusions from charged valence holes expected by many-body scenarios.
\end{abstract}
]
Drastic manifestations are expected from the interaction of a magnetic or charged
impurity with a Fermi sea of electrons\cite{bible-mahan}.
Since the successfull explanation of the divergent
X-ray absorption edges of simple metals\cite{metaux} by
Mahan\cite{mahan} and Nozi\`eres {\it et al.}\cite{noz} thirty years ago,
a Fermi sea of electrons under optical excitation
has been recognized as a model system to study these issues experimentally.
Indeed a many-body
electronic state can develop in presence of the positively charged
electronic vacancy (or core hole) involved in optical processes,
which multiply scatters conduction electrons throughout and along
the Fermi surface. This induces divergences at the Fermi-edge of optical spectra,
the so-called Fermi-edge singularities (FES). FES are
by essence highly sensitive to 
phase space restrictions such as hole localization and
reduced dimensionalities : they cannot form without either
strong hole localization or the dimensionality being unity\cite{noz-dim}.

Electron systems embedded in semiconductor heterostructures
have therefore attracted much attention to test these predictions\cite{gew}.
FES were first observed by Skolnick {\it et al.}\cite{skolnick}
in the low-temperature Photoluminescence (PL) spectrum of
a remotely doped In$_{0.47}$Ga$_{0.53}$As/InP quantum well (QW).
Emphasis was put on the anomalous temperature dependence of the singularity
and on the localization of valence holes by alloy fluctuations.
More recently, experiments by Chen {\it et al.}\cite{chen}
in In$_{0.15}$Ga$_{0.85}$As/AlGaAs QWs with weaker hole
localization also put forward the {\it tunability }of FES, by bringing the first empty
QW subband into resonance with the Fermi level.
It was then proposed that empty conduction subbands could act as
additional scattering channels for Coulomb processes, in qualitative agreement with 
many-body multi-subband numerical calculations
in the infinite hole-mass approximation\cite{hawrylak}.


Although this interpretation is widely referred to, no experiments have been undertaken
to test many-body schemes beyond qualitative agreement. While attempting to do so
with Al$_x$Ga$_{1-x}$As QWs, it appeared to us that {\it non-coulombian }intersubband
scatterings can induce FES as well. The scope of this Letter is to
describe this issue quantitatively. Our experiments demonstrate that
extrinsic processes like alloy-disorder can prevail easily over multiple Coulomb scatterings
of Fermi sea electrons predicted in the framework of many-body scenarios.

The paper is organized as follows. We show that
the formation of FES in Al$_x$Ga$_{1-x}$As QWs is governed by a Fano resonance
mechanism\cite{fano} between Fermi-sea electrons and
discrete excitonic transitions associated with empty conduction subbands.
This model is first validated by the existence of scaling properties of PL spectra
when FES are enhanced by reduced intersubband spacings.
It is then used to gain insight into the microscopic nature of conduction Intersubband
Couplings (ICs) at work. Indeed, the stronger the
ICs, the more divergent the FES, so that one can
probe directly in experiments the efficiency of extrinsic scattering processes such as
alloy-disorder, remote-doping disorder or artificial ICs in lateral
superlattices\cite{melin}.
Experimental results fall in close agreement with microscopic Fano calculations.
We demonstrate that alloy-disorder is the dominant contribution
to observed FES in Al$_x$Ga$_{1-x}$As QWs.

Samples investigated in this work are remotely doped
Al$_{1-x}$Ga$_x$As/Al$_{0.33}$Ga$_{0.67}$As QWs grown on GaAs substrates
by molecular beam epitaxy. They are all of same thickness L$_z$=25nm,
spacers (in the range 5-8 nm) and sheet density $n_s\simeq 8.10^{11}$cm$^{-2}$,
but vary in their aluminium content $x$ ($2.3\%\leq x\leq 7.1\%$).
Confined 2D-levels are represented in fig. 1a. 
The asymmetry of the QW potential
originates in the dipole formed by remote ionized dopants ($z > 0$) and the
Degenerate Electron Gas (DEG)
partially filling the first conduction subband E$_1$.
This confers a PL activity to the two first subbands E$_1$ and E$_2$ with
photocreated valence holes localized on potential fluctuations
at the top of the heavy-hole HH$_1$ subband (fig. 1b).
We perform PL spectroscopy at 1.8K. Samples are optically excited by
a Ti:Sa laser at 1.7 eV, with a $\simeq$1 W.cm$^{-2}$ density so as to
avoid inhomogeneous heating of the DEG within the 40$\mu m^2$ laser spot.
The DEG PL (E$_1$HH$_1$)
extends from lowest wave-vector transitions at E$_g$ up to Fermi wave-vector
transitions at E$_g$+E$_F$ (E$_F \simeq 25$meV).
Without any influence of E$_2$HH$_1$, its oscillator strength decreases monotonously
with energy, because of hole localization and indirect
optical processes\cite{lyo-jones} (see spectrum $\star$ in fig.1c).
The PL of the empty QW subband E$_2$ exhibits a
dominant excitonic feature E$_{2_X}$ of high oscillator
strength and discrete character,
visible either by thermal activation above E$_g$+E$_F$ in PL, or in PL Excitation
spectra.
We define $\Delta$=E$_{2_X}$-E$_g$-E$_F$.
Variations of $\Delta$ are achieved along a given sample by use of the
flux gradients of effusion cells in the epitaxy chamber.
We stop the wafer rotation during the spacer layer growth between dopants
and the QW. The thinner the spacer,
the stronger the electric field at the QW interface.
This tunes the E$_2$-E$_1$ energy separation, while
E$_F$ hardly changes under illumination\cite{note-chen-field}.
As seen from PL spectra of sample A ($x$=7.1$\%$) in fig.1c,
a FES forms and develops\cite{chen} when $\Delta$ is decreased
to zero\cite{magneto-ef}.
\begin{figure}[h]
  \begin{center}
    \leavevmode
    \epsfxsize=0.9\columnwidth
    \epsfbox{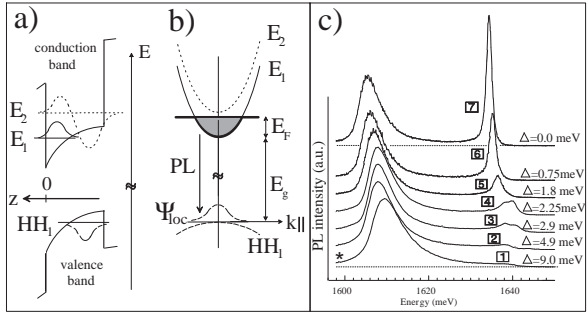}\vskip1mm    
    \caption{a) Schematic 2D-confined conduction and valence envelope-functions
for E$_1$, E$_2$ and HH$_1$. Remote $\delta$-doping is achieved on the $z>0$ side.
b) In-plane $k$-space band-structure showing the PL recombination of Fermi-sea electrons
with photocreated holes at T=0K. Localized HH$_1$ states are sketched by a $k$-space
extended wave-function $\Psi_{loc}$. c) 1.8 K PL spectra of sample A
recorded for various $\Delta$ values.}
\label{figure1}
\end{center}
\end{figure}
To interpret these data, we consider the Fano resonance model depicted in fig.2a.
E$_{2_X}$ is taken as a discrete level coupled with a matrix-element ${\cal W}$
to the continuum of E$_1$HH$_1$ electron-hole pairs.
By assuming an infinite hole mass,
all physical parameters simply refer to the conduction band :
${\cal W}$ equals to the IC between E$_1$ and E$_2$, and the E$_1$HH$_1$ continuum
is populated by the E$_1$ Fermi-Dirac electrons.
We take a parabolic dispersion and a constant PL oscillator strength for E$_1$.
Fano\cite{fano} gave an analytical description of the spreading of a discrete level coupled
to a continuum. Optical FES occur (see fig.2a) due to the partial filling
of E$_1$HH$_1$ near E$_{2_X}$, and get more divergent as $\Delta$ is reduced.
A qualitative agreement with experiments\cite{chen} is thus obtained, without any
particular assumption on the physical nature of ${\cal W}$.

The validation of the Fano resonance model comes from the scaling property
evidenced in fig.2a that all PL spectra for various $\Delta$
should display a common envelope lineshape when plotted
with E$_{2_X}$ as the origin of energies. The comparison with experimental
data is not straightforward because E$_1$-E$_2$ varies experimentally, while
the E$_1$HH$_1$ PL oscillator strength intrinsically depends on energy\cite{lyo-jones}.
We compensate for these nominal E$_1$HH$_1$ PL variations by dividing
all data of fig.1c by the E$_1$HH$_1$ spectrum with
$\Delta \rightarrow \infty$ ($\star$ in fig.1c).
Data are then plotted in fig. 2b with E$_{2_X}$ as
the origin of energies. Remarkably, all spectra superpose on each other in
the range of populated electron states\cite{zonecenter}.
No clear transition exists from convergent to divergent Fermi edges in fig 2b.
This means that FES only appear in raw PL data
when the intrinsic PL decay of E$_1$HH$_1$ at E$_F$ gets balanced
by the positive slope of the Fano envelope for small $\Delta$ values.
\begin{figure}[htbp]
  \begin{center}
    \leavevmode
    \epsfxsize=0.9\columnwidth
    \epsfbox{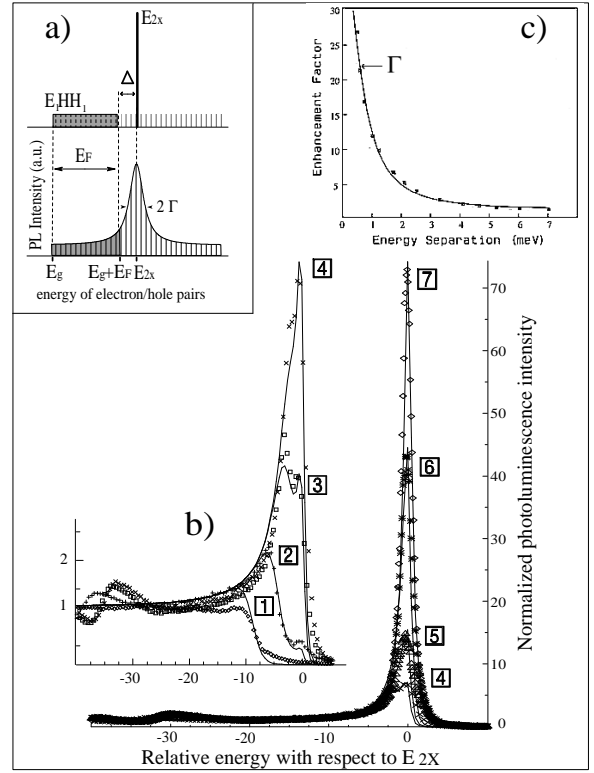}\vskip1mm
    \caption{a) Fano resonance model, where E$_{2_X}$ is seen as a discrete level in
the E$_1$HH$_1$ continuum (top). In the coupled system (bottom),
a FES appears, due to the partial filling of E$_1$ (dark grey).
$2\Gamma$ is the resonance width.
b) Points : data of fig.1c, normalized
by the PL spectrum with $\Delta \rightarrow \infty$ ($\star$
in 1c) and plotted with E$_{2_X}$ as the origin of energies. Lines :
fits using a Fano profile ($q$=12.2 and $\Gamma$=0.60 meV) and
Fermi-Dirac electrons (T=9.0$\pm$1K). c) FES \lq\lq enhancement factor\rq\rq
(points, from ref.[8]) in an In$_{0.15}$Ga$_{0.85}$As QW, fitted (line) by
a Fano profile with $\Gamma$=0.66 meV and $q$=6.6.}
    \label{figure2} \end{center}
\end{figure}
In order to get a quantitative analysis of the envelope profile,
we assume a statistical disorder property for the interaction
${\cal W}$ : $<{\cal W}>$= 0 and
$<{\cal W}^2>\neq 0$. This accounts for alloy disorder, which is
later evidenced to dominate FES in sample A.
The {\it normalized }PL intensity can then be derived analytically
from ref.\cite{fano} :
$$I(E)\ =\ {f(E)}\ .\ {q^2+(E-E_{2_X})^2/\Gamma^2 
\over 1+(E-E_{2_X})^2/\Gamma^2}$$
where E is the PL energy ; $f(E)$ the occupation number in the E$_1$ subband ;
$\Gamma$ is the observed half width at half maximum
of the resonance, only related to the statistical squared interaction average
$<{\cal W}^2>$ and the density of states ${\cal D}$ of the E$_1$ conduction
subband
by $\Gamma =\pi <{\cal W}^2> {\cal D}$ ; $q^2$ is the experimental
oscillator strength of the excitonic resonance relative to the continuum, inversely
proportional to $<{\cal W}^2>$ and ${\cal D}^2$. It depends also strongly
on wavefunction overlaps through the ratio of the
oscillator strengths of E$_2$HH$_1$ and E$_1$HH$_1$. $q^2$ is therefore quite sensitive
to small geometrical fluctuations from sample to sample.

Our data are nicely fitted by this model (fig. 2b),
with Fano parameters $\Gamma$=0.60 meV and $q$=12.2.
All fits use Fermi-Dirac distributions of electrons with an effective temperature
T=9.0$\pm$1K.
This shows that electrons are indeed thermalized, though not with
the lattice at 1.8 K due to the incomplete and slow relaxation of 
photocreated electrons above the E$_2$ subband edge.
The consistence of
small $\Delta$ data with our fit indicates that the E$_{2_X}$ resonance remains
weakly populated enough to stay in the \lq\lq atom-like\rq\rq ~regime\cite{skolnick-ple}.
Also, taking $f(E)$=1 in the Fano lineshape formula, we can check that it fits
the FES \lq\lq enhancement factor\rq\rq~(FES PL intensity divided by its
$\Delta \rightarrow \infty$ value) measured by Chen {\it et al.}
in In$_{0.15}$Ga$_{0.85}$As QW structures\cite{chen}. Their data are indeed
nicely reproduced with $\Gamma$=0.66 meV and $q=6.6$ (fig.2c).

Before quitting the phenomenological level, we focus on the temperature
dependence of FES in systems with explicit intersubband interaction.
Observed thermal quenchings can be
understood simply, even though such quenchings are expected from many-body
theories\cite{ohtaka}, where an actual occupation number
discontinuity is required to enhance multiple Coulomb scatterings at E$_F$.
Here or in ref.\cite{chen}, the FES disappears, only
because the PL relative minimum between E$_g$+E$_F$ and E$_{2_X}$
vanishes with raised temperatures\cite{note-temperature}.

Up to this point, we successfully assessed the model
with respect to $\Delta$ and temperature variations.
We now analyse the microscopic origin of the
Fano parameter $\Gamma$. On the theoretical side, $\Gamma$
only depends on the strength ${\cal W}$ of intersubband
couplings and can be computed for a given microscopic process.
Experimentally, the Fano model predicts that the stronger $\Gamma$, the more divergent
the Fermi-edges at fixed
$\Delta$. By designing appropriate samples, one can thus test : {\it i) }whether
extrinsic scattering processes like alloy-disorder play an effective role on the formation
of FES ; {\it ii) }if experimental variations of $\Gamma$ correlate with
microscopic calculations.

We display in fig.3a the PL spectra of remotely doped quantum wells C, B and A of
with QW aluminium content $x$ equal to 0.023, 0.044 and
0.071 respectively, taken at fixed $\Delta$=4.3$\pm$0.1 meV.
As seen from the global PL lineshapes,
the localization of photocreated valence holes remains constant, dominated by
the roughness at the QW interface (see fig. 1)
rather than by random alloy potential fluctuations. Enhancement
of many-body processes due to hole localization\cite{noz-dim,skolnick}
can therefore be excluded.
Nevertheless, FES get more pronounced with
increased alloy concentration $x$, while the excitonic resonance broadens.
Fitted $\Gamma$ Fano parameters\cite{note-broadening}
linearly increase with $x$ within experimental uncertainty (fig.3b).
This explains the quadratic enhancement of FES visible from fig.3a when $\Gamma$ is
linearly increased in the regime where $\Delta / \Gamma \gg 1$. It also demonstrates that
alloy disorder is the dominant contribution to the IC parameter ${\cal W}$.
\begin{figure}[htbp]
  \begin{center}
    \leavevmode
    \epsfxsize=0.95\columnwidth
    \epsfbox{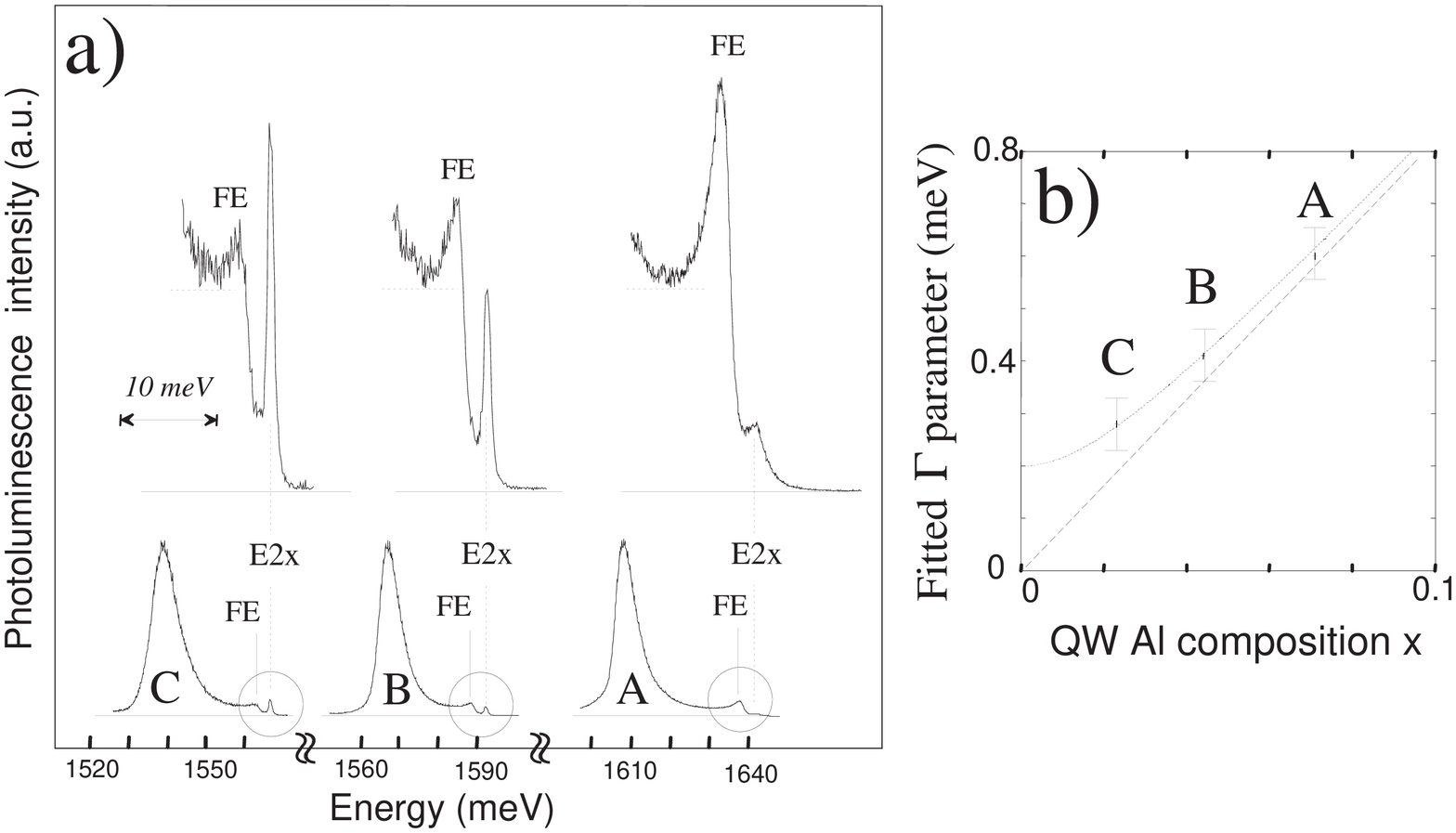}\vskip1mm
    \caption{a) 1.8K PL of samples C, B and A taken at $\Delta$=4.3$\pm$0.1 meV.
The Fermi-edge regions are zoomed in. b) Fitted Fano parameters $\Gamma$
as a function of alloy content.
Microscopic calculation are given without (dotted line) and
with (full line) a residual disorder $\Gamma_o$=0.2 meV (see text).}
    \label{figure3} \end{center}
\end{figure}
To quantify this, we calculate $\Gamma$ microscopically
in the infinite hole-mass approximation\cite{note-gamma-alloy} :
$$\Gamma_{alloy}\ =\ x(1-x)m\Omega_o \delta V^2/L_z\hbar^2$$
This applies for a square
quantum well of width $L_z$, a conduction electron mass $m$, $\delta V$ being the
conduction band offset between pure AlAs and pure GaAs, and $\Omega_o$
the crystal cell volume. With $L_z$=18 nm (representative of the
confinement length of
E$_1$ and E$_2$ wave-functions), $m$=0.07 $m_0$ and $x$=7.1$\%$,
we obtain $\Gamma$=0.61 meV.
This quantitative agreement is striking for such a simple model.
The fit from fig.3b is achieved by introducing a residual scattering $\Gamma_o$=0.2 meV in
the pure GaAs limit and thus taking a Fano parameter
$\Gamma^2={\Gamma_o^2 +\Gamma_{alloy}^2}$. We can also estimate
$\Gamma_{alloy}$$\simeq$0.5 meV for the In$_{0.15}$Ga$_{0.85}$As QWs of ref.\cite{chen},
in close agreement with either our fit (fig.2c) or the empirical two-level
coupling (0.6 meV) measured by Chen {\it et al.}.

We now focus on ICs induced by random positioning of ionized dopants.
Strictly speaking, ${\cal W}$ now depends on the occupation of E$_1$ states.
We nonetheless assume ${\cal W}$ equal to its value for Fermi wave-vector (k$_F$) electrons.
$\Gamma$ can then be computed\cite{fano}, and gets proportional to $\exp{-2k_F . z_o}$ in
the limit of a thick spacer layer $z_o$,
with a prefactor of $\simeq$60 meV for a square QW structure of width L$_z$=20 nm
and a sheet density $n_s\approx 10^{12}$ cm$^{-2}$.
Due to high doping ($k_F\geq 2.10^8$m$^{-1}$),
this predicts a poor efficiency even for very shallow spacers.
This is already visible from fig. 2b where
indirect optical processes\cite{lyo-jones} -~of same physical origin~- only
affect small wave-vector conduction states.
Also, we measured no increase of $\Gamma$ in a sample similar
to C but with a 2.5 nm spacer.

We finally mention the case of FES in tilted lateral superlattices, where
artificial intersubband couplings are created by
a non separable 1D periodic confinement between the growth ($z$) and an in-plane
($x$) direction. This has been shown to promote optical FES\cite{melin}.
In fact, our data also fit to a Fano scheme\cite{tobepublished} with parameter $\Gamma$=3.2 meV.
The larger strength of ICs
compared to 2D QWs explains the formation of pronounced FES even for
large $\Delta$ parameters\cite{melin}. The experimental $\Gamma$ value matches
a microscopic calculation, using a typical value of 30 meV
for the peak-to-peak lateral confinement amplitude\cite{prl-nondopes}.

Microscopic calculations of Fano parameters
fall therefore in quantitative agreement with our experiments, where ICs have been
varied in physical nature and over one decade in amplitude.
This demonstrates that non-coulombian scattering processes
can efficiently control the formation of FES in multi-subband semiconductor structures,
in the simple picture of a band-structure partially filled by a DEG.
We stress that Coulomb
processes actually depend on the accurate distribution of charged particles
with respect to empty subbands and {\it do not }quantitatively match
Fano lineshapes\cite{hawrylak},
even if a qualitative analogy was
underlined in early many-body interpretations\cite{chen,hawrylak}.
The question of extrinsic ICs can also be raised about
ref.\cite{skolnick}. A FES enhancement of a factor $\simeq$2
would correspond to $q\simeq \Delta / \Gamma$ for a Fano process involving empty subbands.
Estimating $\Delta \simeq$30 meV\cite{skolnick} and
$\Gamma_{alloy}=$1.0 meV, this criterium discards random alloy-disorder processes provided
$q\ll 30$. An evaluation of actual disorder processes
and relevant absorption data on E$_{2_X}$
are definitely lacking to give any conclusive answer, but the QW doping on both sides
used in ref.\cite{skolnick} should reduce advantageously the
E$_2$ and HH$_1$ overlap and thus $q$. Bringing $q$-values down by optical selection rules
indeed suppresses excitonic enhancement effects\cite{hawrylak}.

In conclusion, this paper shows that extrinsic intersubband scatterings
must be carefully evaluated while analysing FES.
This applies especially to lower
dimensionality issues\cite{noz-dim} since disorder gets generally enhanced,
and conduction subband spacings lowered.
Realistic many-body theories should thus include disorder and
multiple subbands.
Our work should extend under magnetic field where the
interaction of excitons and a DEG is debated\cite{prl-comment}.
Extremely low-disorder structures seem required for
clear electron-electron manifestations\cite{gravier}.

We thank F. Petit, P. Denk, F. Lelarge, A. Cavanna, R. Planel,
C. Tanguy, B. Jusserand and B. Etienne for stimulating discussions.
This work was supported by DRET and
a SESAME grant of the R\'egion Ile de France.


\begin{thebibliography}{99}
\bibitem{magneto-ef} Reported $\Delta$ values stem from the fit
in fig.2b (see text). They fall in close agreement with magneto-optical PL data,
which provide an actual measurement of $E_g+E_F$.
\bibitem{zonecenter} Small discrepancies occur around the low-energy peak of the DEG PL,
all the more pronounced with thinner spacer layers. This signs the increased
efficiency of indirect optical processes for less
remote dopants, while they are taken constant in the normalization procedure.
\end{thebibliography}
\end{document}